\documentclass[11pt]{article}

\usepackage[text={17cm,24cm}]{geometry}

\RequirePackage[OT1]{fontenc}
\RequirePackage{amsthm,amsmath}
\RequirePackage[numbers,square]{natbib}
\RequirePackage{hyperref}

\usepackage{mathptmx}

\usepackage{helvet}
\usepackage{courier}
\usepackage{type1cm}

\usepackage{graphicx}        
\usepackage[bottom]{footmisc}

\usepackage{latexsym,amssymb}
\usepackage{amsmath}
\usepackage{subfig}
\usepackage{longtable}
\usepackage{siunitx}
\usepackage{afterpage}
\usepackage{morefloats}
\usepackage[section]{placeins}

\usepackage[shortlabels]{enumitem}
\newlist{arguments}{description}{1}
\setlist[arguments]{style=sameline}
\newlist{inlinelist}{enumerate*}{1}
\setlist*[inlinelist,1]{label=(\alph*), itemjoin={{, }}, itemjoin*={{, and }}}

\usepackage{algpseudocode}
\usepackage{capt-of}

\algnewcommand\algorithmicinput{\textit{Input:}}
\algnewcommand\Input{\item[\algorithmicinput]}
\algnewcommand\algorithmicoutput{\textit{Output:}}
\algnewcommand\Output{\item[\algorithmicoutput]}

\newcounter{ssaalg}[section]
\makeatletter
\renewcommand\thessaalg{\thesection.\@arabic\c@ssaalg}%
\def\ext@ssaalg{loa}
\def\fnum@ssaalg{\textsc{Algorithm~\thessaalg}}
\makeatother
\newenvironment{algorithm}[2]{%
    \begingroup
    \captionof{ssaalg}{#1\label{#2}}\nopagebreak%
    \begin{algorithmic}[1]%

    \addtolength{\itemsep}{2pt}}{%
    \end{algorithmic}%
    \vskip 1em%
    \endgroup}

\graphicspath{{img_ch1/}}

\usepackage{graphics}
\usepackage{tikz}
\usepackage{ifpdf}
\ifpdf\usepackage{epstopdf}\fi
\usepackage{url}

\usepackage{amsbsy}

\usepackage{fancyvrb}
\DefineVerbatimEnvironment{Code}{Verbatim}{fontsize=\small,fontshape=sl}
\DefineVerbatimEnvironment{Func}{Verbatim}{fontsize=\small,fontshape=tt}
\makeatletter
\newcommand\code{\bgroup\@makeother\_\@makeother\~\@makeother\$\@codex}
\def\@codex#1{{\normalsize\ttfamily\hyphenchar\font=-1 #1}\egroup}
\makeatother
\newcommand{\pkg}[1]{\textsc{#1}}
\let\proglang=\textsf

\makeatletter
\def\verbatim@font{\small\ttfamily}
\makeatother

\usepackage{euscript}

\newcommand{\cM}{\EuScript{M}}

\newcommand{\cT}{\EuScript{T}}

\newcommand{\rmT}{\mathrm{T}}

\newcommand{\tS}{\mathbb{S}}

\newcommand{\tX}{\mathbb{X}}
\newcommand{\tY}{\mathbb{Y}}

\newcommand{\bfA}{\mathbf{A}}
\newcommand{\bfB}{\mathbf{B}}
\newcommand{\bfC}{\mathbf{C}}

\newcommand{\bfP}{\mathbf{P}}
\newcommand{\bfQ}{\mathbf{Q}}

\newcommand{\bfX}{\mathbf{X}}
\newcommand{\bfY}{\mathbf{Y}}

\newcommand{\calC}{\mathcal{C}}

\newcommand{\calH}{\mathcal{H}}

\newcommand{\calR}{\mathcal{R}}

\newcommand{\bt}{\begin{theorem}}
\newcommand{\et}{\end{theorem}}
\newcommand{\bl}{\begin{lemma}}
\newcommand{\el}{\end{lemma}}
\newcommand{\bp}{\begin{proposition}}
\newcommand{\ep}{\end{proposition}}
\newcommand{\bc}{\begin{corollary}}
\newcommand{\ec}{\end{corollary}}

\newcommand{\bd}{\begin{definition}\rm}
\newcommand{\ed}{\end{definition}}
\newcommand{\bex}{\begin{example}\rm}
\newcommand{\eex}{\end{example}}
\newcommand{\br}{\begin{remark}\rm}
\newcommand{\er}{\end{remark}}

\newcommand{\btbh}{\begin{table}[!htbp]}
\newcommand{\etb}{\end{table}}
\newcommand{\bfgh}{\begin{figure}[!htbp]}
\newcommand{\efg}{\end{figure}}

\newcommand{\bea}{\begin{eqnarray*}}
\newcommand{\eea}{\end{eqnarray*}}
\newcommand{\be}{\begin{eqnarray}}
\newcommand{\ee}{\end{eqnarray}}

\newcommand{\suml}{\sum\limits}

\newcommand{\ve}{\varepsilon}

\newcommand{\lm}{\lambda}
\def\wtilde{\widetilde}

\newcommand{\ra}{\rightarrow}

\def\bproof{\textit{Proof.\ }}
\def\eproof{\hfill$\Box$\smallskip}

\def\spaceR{\mathsf{R}}

\def\sspan{\mathop{\mathrm{span}}}
\def\rank{\mathop{\mathrm{rank}}}

\makeatletter
\def\adots{\mathinner{\mkern2mu\raise\p@\hbox{.}
\mkern2mu\raise4\p@\hbox{.}\mkern1mu
\raise7\p@\vbox{\kern7\p@\hbox{.}}\mkern1mu}}
\newcommand{\l@abcd}[2]{\hbox to\textwidth{#1\dotfill #2}}
\makeatother

\newtheorem{proposition}{Proposition}
\newtheorem{corollary}{Corollary}
\newtheorem{theorem}{Theorem}
\newtheorem{remark}{Remark}
\newtheorem{lemma}{Lemma}
\newtheorem{definition}{Definition}

\begin{document}

\title{Semi-nonparametric singular spectrum analysis with projection}

\author{Nina Golyandina\footnote{nina@gistatgroup.com},  Alex Shlemov\footnote{shlemovalex@gmail.com}}
\date{St.Petersburg State University}
\maketitle

\begin{abstract}
Singular spectrum analysis (SSA) is considered for decomposition of time
series into identifiable components. The Basic SSA method is nonparametric and constructs
an adaptive expansion based on singular value decomposition. The investigated
modification is able to take into consideration a structure given in advance
and therefore can be called semi-nonparametric. The approach called SSA with projection
includes preliminary projections of rows and columns of the series’ trajectory
matrix to given subspaces. One application of SSA with projection is the extraction
of polynomial trends, e.g., a linear trend. It is shown that SSA with projection can
extract polynomial trends much better than Basic SSA, especially for linear trends. Numerical examples including comparison with the least-square
approach to polynomial regression are presented.
\end{abstract}

\section{Introduction}
Singular spectrum analysis (SSA) can solve a wide range of problems in the time series analysis, from
the series decomposition on the interpretable series components to forecasting, missing data imputation,
parameter estimation and many others, see, e.g., \cite{Vautard.etal1992,Elsner.Tsonis1996,Golyandina.etal2001,Golyandina.Zhigljavsky2012} and references within.
The key feature of SSA is that the basic method is model-free, does not need a-priori information and
therefore constructs an adaptive decomposition of a time series into a sum of e.g. a non-parametric trend,
periodic components and noise (see \cite{Florinsky.etal2009,Vityazev.etal2010,Alexandrov.etal2012,Shaharudin.etal2015,Unnikrishnan.Jothiprakash2015} among others for application of SSA to the problem of trend extraction).
This can be considered as a great advantage of the SSA-family methods for comparison with parametric ones.
However, sometimes there is a-priori information about the
considered time series. For example, the trend can be expected as linear or polynomial.

In \cite[Section 1.7.1]{Golyandina.etal2001}, SSA with single and double centering is developed to extract
constant or linear trends with better accuracy.
We generalize this approach. Approaches, which deal with a combination of parametric and nonparametric
models, are sometimes called semi-parametric if the parametric part of the model is of interest and semi-nonparametric
if both parts are important, see the references \cite{Chen2007} and \cite{Ichimura.Todd2007}
as examples of such approaches to statistical econometric  problems.

\smallskip
Let us explain the motivation for the suggested approach, which can be considered as a semi-nonparametric variation
of singular spectrum analysis.

In SSA, the separability theory is responsible for a proper decomposition and component extraction.
The separability of a series component means that the method is able to extract this time series component
from the observed series, which is a sum of many components.
Basic SSA is able to approximately separate a trend (e.g., a linear trend) from oscillations. However,
there is no series,
which can be exactly separated from a linear trend. As a consequence, the separation accuracy is not high.
It is shown in \cite[Sections 1.7.1 and 6.3.2]{Golyandina.etal2001} that SSA with double centering weakens the separability conditions
and therefore improves the accuracy in conditions of approximate separability. Thus, it is expected that, within the SSA-family methods,
SSA with projection can improve separability for components of a specific structure, which is in accordance with the projection
subspaces.

In comparison with the linear regression technique (we will further mean the least squares approach to the estimation of regression parameters), SSA with double centering differs by the statement of the problem.
Linear regression minimizes the prediction error, while SSA tries to separate the series components themselves
using their orthogonality.
For example, for a series with common term $x_n=t_n+s_n$, where $t_n=an+b$ and $s_n=A \sin(2\pi \omega n +\phi)$,
the least-squares approach generally cannot estimate the linear trend $t_n$ with no error, while in the conditions of separability
SSA with double centering is able to find the exact linear trend.
For long time series, the linear regression and SSA yield close estimates of the linear trend.
 Note that for the case of approximate separability
the trend found by SSA with double centering will be only close to a straight line, while the linear regression always provides
a linear function as a trend estimation.
The analogous relation between the parametric regression and SSA with projection is expected for the general case of
polynomial trends. In particular,  we can suppose that for time series with seasonality the `SSA with projection' method
will be able to extract linear and polynomial trends more accurately than the parametric regression approach.
It is important that the use of projection on a fixed basis does not contradict the non-parametric nature of SSA.
Moreover, if the basis is chosen incorrectly, the decomposition will be not optimal and the trend estimate will be less accurate;
however, the estimate will not have a considerable bias, since it can be accomplished by components of the adaptive part
of the whole decomposition. This is not the case for the parametric approach.

The Basic SSA method consists of trajectory matrix construction from the original time series, its decomposition into a sum of
rank-one matrices by SVD,
their grouping and then each group's return to time series to obtain a decomposition of the original time series
into a sum of identifiable components. The grouping of the SVD can be considered as a projection of
the trajectory matrix columns on a subspace,
which is adaptively constructed based on the distinguished features of SVD decomposition.
SSA with projection starts with projections of trajectory matrix columns and rows
on subspaces chosen in advance and then decomposition of the residual, by the same way as in Basic SSA.
In particular, SSA with double centering uses the projections on the subspaces spanning the vectors
with elements equal to $1$.
A natural application of SSA with projection, which is mostly considered in this paper, serves for extraction of polynomial trends;
however, the suggested method can be applied to a wider range of problems, e.g., for the use
of information about a supporting series.

\smallskip
The structure of the paper is as follows.
We start with a short description of the algorithm of Basic SSA and standard separability notion (Section~\ref{sec:BasicSSA}).
Section~\ref{sec:SSAproj} is devoted to generalizing centering
used in SSA and contains  the underlying theory, including the proof of the algorithm and the separability conditions.
Section~\ref{sec:ex} demonstrates the examples of the algorithm application for trend extraction.
The real-life examples are studied in Sections~\ref{sec:gasoline} and \ref{sec:co2} to show the relation between
Basic SSA, SSA with projection and the linear regression (least-squares) approach.
Numerical comparison is performed in Section~\ref{sec:comparison}.
The paper is summarized and conclusions are drawn in Section~\ref{sec:concl}.

\section{Necessary information}
\label{sec:BasicSSA}
\subsection{Algorithm of Basic SSA}

Consider a real-valued time series $\tX=\tX_N=(x_1,\ldots,x_{N})$ of length $N$.
Let $L$ ($1<L<N$) be some integer called {\em window length} and $K=N-L+1$.

For convenience, denote $\cM_{L,K}$ the space of matrices of size $L\times K$ and
$\cM_{L,K}^{(H)}$ the space of Hankel matrices of size $L\times K$. Consider
the {\em lagged vectors}
$X_i=(x_{i},\ldots,x_{i+L-1})^\rmT$, $i=1,\ldots,K$,
and the {\em trajectory matrix} $\bfX=[X_1:\ldots:X_K]\in \cM_{L,K}^{(H)}$ of the series $\tX_N$.

Define the one-to-one embedding operator $\cT: \spaceR^{N} \mapsto \cM_{L,K}^{(H)}$.
as $\cT(\tX_N)=\bfX$.
Also introduce the projector $\calH$ (in Frobenius norm) of
 $\cM_{L,K}$ to $\cM_{L,K}^{(H)}$. Projection is performed by the change of
 entries on auxiliary
diagonals $i+j=\mathrm{const}$ to their averages along the diagonal.

The Basic SSA algorithm consists of four steps.

\smallskip
{\bf 1st step: Embedding.}
Let $L$ be chosen. At this step the $L$-trajectory matrix is composed:
 $\bfX=\cT(\tX_N)$.

\smallskip
{\bf 2nd step: Singular Value Decomposition (SVD).}
The SVD of the trajectory matrix is constructed:
\be
\label{eq:elem_matr}
\bfX=\sum_{i=1}^d \sqrt{\lambda_i}U_i V_i^\rmT=\bfX_1 + \ldots + \bfX_d,
\ee
where $\sqrt{\lambda_i}$ are singular values,
$U_i$ and $V_i$ are left and right singular vectors of $\bfX$,
$\lambda_1\geq\ldots\geq \lambda_d > 0$, $d=\rank(\bfX)$.

The triple $(\sqrt{\lm_i},U_i,V_i)$ is called $i$th {\em
eigentriple} (abbreviated as ET).

\smallskip
{\bf 3rd step: Eigentriple grouping.}
The grouping procedure
partitions the set of indices $\{1,\ldots,d\}$ into $m$ disjoint subsets
$I_1,\ldots,I_m$.

Define $\bfX_I=\sum_{i\in I} \bfX_i$.
The expansion \eqref{eq:elem_matr} leads to the decomposition
\be
\label{eq:mexp_g}
\bfX=\bfX_{I_1}+\ldots+\bfX_{I_m}.
\ee

If $m=d$ and $I_j=\{j\}$,
$j=1,\ldots,d$, then the corresponding grouping is called \emph{elementary}.

\smallskip
{\bf 4th step: Diagonal averaging.}
Obtain the series by diagonal averaging of the matrix components of \eqref{eq:mexp_g}:
$\wtilde\tX^{(k)}_N = \cT^{-1} \calH \bfX_{I_k}$.

\smallskip
Thus, the algorithm results in the constructed decomposition of the observed time series
\be
\label{eq:sexp_f}
  \tX_N = \suml_{k=1}^m\wtilde\tX^{(k)}_N.
\ee

A typical example of  \eqref{eq:sexp_f} is the decomposition into a sum of a trend, oscillations and noise.

\begin{remark}
\label{rem:svd_proj}
Columns of a grouped matrix $\bfX_I$ are the projections of columns of the trajectory matrix $\bfX$
to $\sspan(U_i,\, i\in I)$. Rows of $\bfX_I$ are the projections of rows of $\bfX$ to $\sspan(V_i,\, i\in I)$.
\end{remark}

\subsection{Separability by Basic SSA}
\label{sec:sep}
To understand how SSA works, the notion of separability is very important.
Separability of two time series $\tX^{(1)}_N$ and $\tX^{(2)}_N$ signifies the possibility of extracting
$\tX^{(1)}_N$ from the observed sum $\tX_N=\tX^{(1)}_N + \tX^{(2)}_N$.
This means that there exists a grouping at Grouping step such that
$\wtilde\tX^{(k)}_N=\tX^{(k)}_N$.

By properties of the SVD, the separability is concluded in the orthogonality of the column and row spaces
of the trajectory matrices of the series $\tX^{(1)}_N$ and $\tX^{(2)}_N$.
In the case of approximate (asymptotic) separability $\wtilde\tX^{(k)}_N\approx \tX^{(k)}_N$
we obtain the condition of approximate (asymptotic) orthogonality.

For sufficiently long time series, SSA can approximately separate, for example, a signal and noise, sine waves with different
frequencies, a trend and a seasonality \citep{Golyandina.etal2001,Golyandina.Zhigljavsky2012}.

The introduced separability, which is called weak separability, means that at the SVD step there exists
such an SVD that allows the proper grouping. Strong separability means that each SVD decomposition
allows the proper grouping.
Several nonparametric modifications of SSA for improvement of the weak and strong separability
are considered in \cite{Golyandina.Shlemov2014}.
In this paper we will improve the separability by a semi-nonparametric variation.

\subsection{Series of finite rank and series governed by linear recurrence relations}
\label{sec:lrr}
Let us describe the class of series of finite rank, which is natural for SSA.
In particular, only such time series can be exactly separated by Basic SSA.

Define the $L$-rank of a series $\tX_N$ as the rank of its $L$-trajectory matrix.
Series with rank-deficient trajectory matrices are of special interest.
A time series is called \emph{time series of finite rank} $r$ if its
$L$-trajectory matrix has rank $r$ for any $L\ge r$ (it is convenient to
assume that $L\le K$).
We will call the column and row spaces of the trajectory matrices
column and row spaces of the series respectively.

Under some unrestrictive conditions \citep[Section 5.2]{Golyandina.etal2001}, series $\tS_N$ of finite rank $r$
is governed by a linear recurrence relation (LRR) of order $r$, that is,
\be
\label{eq:lrf}
s_{i+r}=\sum_{k=1}^r a_k s_{i+r-k},\ 1\leq i\leq N-r,\ a_r\neq 0.
\ee

The LRR~\eqref{eq:lrf} is called minimal and $r$ is called the dimension of the series. Let us describe how
we can restore the form of the time series by means of the minimal LRR.

\begin{definition}
    The polynomial $P_r(\mu)=\mu^r - \sum_{k=1}^r a_k \mu^{r-k}$ is called a
    {characteristic polynomial} of the LRR \eqref{eq:lrf}.
\end{definition}
Let the time series $\tS_{\infty}=(s_1,\ldots,s_n,\ldots)$ satisfy the LRR
\eqref{eq:lrf} with $a_r\neq 0$ and $i\geq 1$. Consider the characteristic
polynomial of the LRR \eqref{eq:lrf} and denote its different (complex) roots by
$\mu_1,\ldots,\mu_p$, where $p \leq r$.  All these roots are non-zero as
$a_r\neq 0$.  Let the multiplicity of the root $\mu_m$ be $k_m$, where $1\leq
m\leq p$ and $k_1+\ldots+k_p=r$.
We will call $\mu_j$ \emph{characteristic roots} of the series governed by an LRR.

It is well-known that the time series $\tS_{\infty}=(s_1,\ldots,s_n,\ldots)$
satisfies the LRR $\eqref{eq:lrf}$ for all $i\ge 0$ if and only if
\be
\label{eq:GEN_REQ}
s_n = \suml_{m=1}^p \left(\suml_{j=0}^{k_m-1} c_{mj} n^j\right) \mu_m^n,
\ee
where the coefficients $c_{mj}$ are determined by the first $r$ series terms.
For real-valued time series, \eqref{eq:GEN_REQ} implies
that the class of time series governed by the LRRs consists of a sum of products
of polynomials, exponentials and sinusoids.

Rank of the series is equal to the number of non-zero terms in
\eqref{eq:GEN_REQ}.  For example, an exponentially-modulated sinusoid
$s_n=A e^{\alpha n}\sin(2\pi \omega n +\phi)$ is constructed from two
conjugate complex roots $\mu_{1,2}=e^{\alpha\pm \textsl{i} 2\pi \omega} = \rho
e^{\pm \textsl{i} 2\pi \omega}$ if its frequency
$\omega\in(0,0.5)$. Therefore, the rank of this exponentially-modulated sinusoid
is equal to $2$.  The rank of an exponential is equal to $1$, the rank of a linear
function corresponding to the root $1$ of multiplicity $2$ equals $2$, and so on.

Also, the representation \eqref{eq:GEN_REQ} helps to easily construct the bases of trajectory spaces of complex time
series governed by LRRs: they are constructed from the linearly independent vectors
$\left(0^j \mu_m^0, 1^j \mu_m^1, \ldots, (L-1)^j \mu_m^{L-1}\right)^\rmT$.
For linear series, the basis consists of $(1,\ldots,1)^\rmT$ and $(0,1,2,\ldots, L-1)^\rmT$.

\section{SSA with projection}
\label{sec:SSAproj}

Let us consider a time series $\tX$ of length $N$, a window length $L$,
$K=N-L+1$, the trajectory matrix $\bfX$ of the series $\tX$.

A general form of the considered modification can be expressed as
\begin{itemize}
    \item Calculation of a special matrix $\bfC=\bfC_\bfX$ based on a-priori information.
    \item Computation of $\bfX'=\bfX-\bfC$.
    \item Construction of the SVD: $\bfX'=\sum_{i=1}^{d'} \sqrt{\lambda'_i} U'_i (V'_i)^\rmT$.
\end{itemize}

Thus, we have the decomposition $\bfX=\bfC+\sum_{i=1}^{d'} \sqrt{\lambda'_i}
U'_i (V'_i)^\rmT$.

Centering, which is a particular case of the general scheme, is considered
in the following forms \citep{Golyandina.etal2001}:
\begin{enumerate}
    \item \emph{Single row centering} when $\bfC$ corresponds to averaging by rows, that is, each element of a row of $\bfC$ consists of
    the average of the corresponding row of the trajectory matrix.
    \item \emph{Single column centering} when $\bfC$ corresponds to averaging by
    columns.
    \item \emph{Double centering} when $\bfC$  corresponds to  averaging by both rows and
    columns.
\end{enumerate}

Single centering can be considered as a projection of rows or columns of $\bfX$ on $\sspan(E_M)$, where
$E_M=(1,\ldots,1)^\rmT\in \spaceR^M$, $M$ is equal to $L$ or $K$. Therefore,
centering in SSA can be considered as a preliminary projection of the trajectory
matrix on a given subspace; the residual matrix  $\bfX'$ will be subsequently expanded by SVD or any other decomposition.

Let us generalize this approach to projections to arbitrary spaces.  Denote a
basis of the column projection space $(P_i, i=1,\ldots, p)$ and/or a basis
of the row projection space $(Q_i, i=1,\ldots, q)$.
Let $\Pi_\mathrm{col}:\, \spaceR^L \ra \sspan(P_i, i=1,\ldots, p)$ and
$\Pi_\mathrm{row}:\, \spaceR^K \ra \sspan(Q_i, i=1,\ldots, q)$ be orthogonal
projectors.
For any $\bfY \in \cM_{L,t}$, denote $\Pi_\mathrm{col}(\bfY)$ the matrix consisting of the
columns, which result from projections of the columns of $\bfY$, while
for any $\bfY \in \cM_{t,K}$ denote $\Pi_\mathrm{row}(\bfY)$ the matrix consisting of the
rows, which result from projections of the rows of $\bfY$.

In SSA with projection,
the scheme of SSA with centering is extended to arbitrary projections, that is,
$\bfC=\Pi_\mathrm{col}(\bfX)$ for column projection,
$\bfC=\Pi_\mathrm{row}(\bfX)$ for row projection and
$\bfC=\Pi_\mathrm{both}(\bfX)$ for double projection, where
$\Pi_\mathrm{both}(\bfX)=\Pi_\mathrm{row} (\bfX) + \Pi_\mathrm{col}
(\bfX-\Pi_\mathrm{row} (\bfX))$. If either the column or row basis is absent (that is,
the corresponding projection should not be performed), then we
formally set the corresponding projector to be the zero operator implying
$\bfC=\Pi_\mathrm{both}(\bfX)$ for any mode.

Note that the method of SSA with projection differs from Basic SSA only in the Decomposition step:
\be
\label{eq:ssaproj}
\bfX=\bfC+\sum_{i=1}^{d'} \sqrt{\lambda'_i} U'_i (V'_i)^\rmT,
\ee
where $\sum_{i=1}^{d'} \sqrt{\lambda'_i} U'_i (V'_i)^\rmT$ is the SVD of $\bfX-\bfC$. Let us show that
\eqref{eq:ssaproj} can be represented as a sum of elementary matrices
and therefore Reconstruction steps can be performed in the same way
as done in Basic SSA.

Without loss of generality we assume that $\{P_i, i=1,\ldots, p\}$ and $\{Q_i, i=1,\ldots, q\}$
are orthonormal systems (otherwise, we can perform ortho-normalization).
Denote $\bfP=[P_1:\ldots:P_p]$, $\bfQ=[Q_1:\ldots:Q_q]$.
Then $\Pi_\mathrm{col}(\bfY)=\bfP\bfP^\rmT\, \bfY = \sum_{i=1}^{p} P_i (\bfY^\rmT P_i)^\rmT$
and $\Pi_\mathrm{row}(\bfY)= \bfY\, \bfQ \bfQ^\rmT = \sum_{i=1}^{q} (\bfY Q_i) Q_i^\rmT$.
Since $\bfC=\Pi_\mathrm{both}(\bfX)$ and can be expressed as a sequential application
of the projection operators $\Pi_\mathrm{row}$ and
$\Pi_\mathrm{col}$,
\eqref{eq:ssaproj}
is  a  decomposition of $\bfX$ on elementary matrix components unambiguously defined.
For double projection, this representation depends on the order
of projections; we will apply the row projector first.

Thus, the matrix
$\bfC$ can be considered as a sum of $p + q$ elementary matrices of the forms
$\sigma^{(c)}_i P_i \wtilde{Q}_i^\rmT$, $i=1,\ldots,p$,
and $\sigma^{(r)}_i \wtilde{P}_i Q_i^\rmT$, $i=1,\ldots,q$.
The triples $(\sigma^{(c)}_i , P_i ,\wtilde{Q}_i)$ and
$(\sigma^{(r)}_i, \wtilde{P}_i ,Q_i)$ have the same meaning as
eigentriples.

The Reconstruction stage is exactly the same as in the Basic
SSA method.  Note that it makes little sense to include the eigentriples
produced by projections to different groups, since the projections
are performed on the subspaces as a whole.

\subsection{Appropriate class of time series}

For SSA with projection, a known series component with a trajectory
matrix $\bfY$ should be in agreement with projection so that $\Pi_\mathrm{col}(\bfY)=\bfY$
for column projection, $\Pi_\mathrm{row}(\bfY)=\bfY$ for row projection and
$\Pi_\mathrm{both}(\bfY)=\bfY$ for double projection.

Clearly, for column and row projections, this is true if the  corresponding
projection is performed on the column or row trajectory space of the known series component.
For example, the trajectory space of an exponential component $s_n=\mu^n$
spans $(1,\mu,\ldots,\mu^{L})^\rmT$, while the trajectory space of a linear function
$s_n=an+b$ spans $(1,1,\ldots,1)$ and $(1,2,\ldots,L)^\rmT$
for any $b$ and non-zero $a$.

Let us derive a condition sufficient for  $\Pi_\mathrm{both}(\bfX)=\bfX$ to hold for the general case of the double projection.

\begin{lemma}
 \label{proj_matrix}
   Let $\Pi_\mathrm{row}(\bfQ^\rmT)=\bfQ^\rmT$, $\Pi_\mathrm{col}(\bfP)=\bfP$ for $\bfP\in \cM_{L,p}$
   and $\bfQ\in \cM_{L,q}$. Then $\Pi_\mathrm{both}(\bfX)=\bfX$ for
    \begin{equation}
        \label{eq:dec2mat}
        \bfX=\wtilde{\bfP}{\bfQ}^{\rmT}+{\bfP}\wtilde{\bfQ}^{\rmT},
    \end{equation}
    where $\wtilde{\bfP}\in \cM_{L,p}$ and $\wtilde{\bfQ}\in \cM_{K,p}$.
\end{lemma}

\bproof
    By the assumption, $\Pi_\mathrm{row}(\bfA\bfQ^\rmT)=\bfA\bfQ^\rmT$
    for any $t$ and matrix $\bfA \in \cM_{t,q}$, while $\Pi_\mathrm{col}(\bfP\bfB^\rmT)=\bfP\bfB^\rmT$
    for any matrix $\bfB \in \cM_{t,q}$.  Therefore,
    \begin{multline*}
        \Pi_\mathrm{both}\bfX = \wtilde{\bfP}{\bfQ}^{\rmT} + \Pi_\mathrm{row} ({\bfP}\wtilde{\bfQ}^{\rmT})+
        {\bfP}\wtilde{\bfQ}^{\rmT} + \Pi_\mathrm{col} (\wtilde{\bfP}{\bfQ}^{\rmT})\\ -
        \Pi_\mathrm{col}( \Pi_\mathrm{row}(\wtilde{\bfP}{\bfQ}^{\rmT}+{\bfP}\wtilde{\bfQ}^{\rmT}))=\bfX,
    \end{multline*}
    since $\Pi_\mathrm{col}\circ \Pi_\mathrm{row} \equiv \Pi_\mathrm{row} \circ
    \Pi_\mathrm{col}$.
\eproof

It is easy to check that the trajectory matrix of a linear series
satisfies the conditions of Lemma~\ref{proj_matrix} for the case
of double centering. However, for a general case the approach based
on characteristic roots is more convenient. We start with a technical lemma.

\begin{lemma}
    \label{lem:poly_dec}
    For any polynomial $P_d$ of order $d$ and for any $m$ and $l$ such that $m +
    l = d - 1$ the following expansion can be constructed:
    \begin{equation*}
        P_d(i + j) = P_{m, d}(i, j) + P_{d, l}(i, j),
    \end{equation*}
    where $P_{u, v}(i, j)$ denotes a polynomial of $i$ and $j$ of order $(u, v)$.
\end{lemma}
\bproof
    This lemma is proved by an appropriate grouping of the monomials $C_{p,q} i^p j^q$, $p + q \leq d$, of $P_d(i + j)$.
\eproof

Recall that a series governed by an LRR, whose characteristic polynomial
has the given set of roots called characteristic roots, is of the form
\eqref{eq:GEN_REQ}.

\begin{theorem}
    Let a series $\tY^{(m)}$ ($m=1,2$) be governed by an LRR of order
    $r_m$, $\bfY^{(m)}$ be its trajectory matrix. Let $\{\mu_j; \;j=1,\ldots,s\}$ be the set containing the     characteristic roots
    of both series. Assume that $\tY^{(m)}$ has roots $\mu_j$, $j=1,\ldots,s$,
    with multiplicities $d_j^{(m)}\ge0$, $\sum_{j=1}^s d_j^{(m)}=r_m$.  Let
    $\Pi_\mathrm{col}$ be the projector on the column space $\calC$ of
    $\bfY^{(1)}$, $\Pi_\mathrm{row}$ be the projector on the row space $\calR$
    of $\bfY^{(2)}$, $\Pi_\mathrm{both} = \Pi_\mathrm{col} + \Pi_\mathrm{row} -
    \Pi_\mathrm{col}\circ\Pi_\mathrm{row}$.  Then $\Pi_\mathrm{both}(\bfX) = \bfX$ if
    and only if the set of characteristic roots of the series $\tX$ consists of
    the roots $\mu_j$, $j=1,\ldots,s$, of multiplicities $d_j\le d_j^{(1)} +
    d_j^{(2)}$.
\end{theorem}

\bproof
    Due to linearity of projectors and linear dependence of $\Pi_\mathrm{both}$
    on $\Pi_\mathrm{row}$ and $\Pi_\mathrm{col}$, it is sufficient to prove the
    theorem for the case of one root $\mu$.  Let $\tY^{(1)}$ have the
    characteristic root $\mu$ of multiplicity $p$, $\tY^{(2)}$ have the
    characteristic root $\mu$ of multiplicity $q$.

    Thus, we should prove that $\Pi_\mathrm{both}(\bfX) = \bfX$ if and only if
    the series $\tX$ has the form $x_k = P_{t}(k)\mu^k$, where $t\leq p+q-1$.
    It is sufficient to take $t=p+q-1$.

    By Lemma~\ref{lem:poly_dec}
    \begin{multline*}
        P_{p + q - 1}(i + j)\mu^i \mu^j\\ =
        P_{p - 1, p + q - 1}(i, j)\mu^i \mu^j + P_{p + q -1, q - 1}(i, j)\mu^i \mu^j.
    \end{multline*}
    This means that \eqref{eq:dec2mat} holds for $\bfQ\in  \cM_{K,q}$
    and $\bfP\in \cM_{L,p}$ such that
    the column space of $\bfQ$ coincides with $\calR$ and the column space of
    $\bfP$ coincides with $\calC$.

    Since the dimension of the space of trajectory matrices that are kept by
    the projector $\Pi_\mathrm{both}$ is equal to $r=r_1+r_2$, we found all
    such matrices. This completes the proof.
\eproof

\begin{corollary}
    \label{cor:double_mult}
    Let $\tY$ be a series of dimension $r$, $\bfY$ be its trajectory matrix,
    $\Pi_\mathrm{row}$ be the projection on its row trajectory space,
    $\Pi_\mathrm{col}$ be the projection on its column trajectory space.
    Consider the series $\tX$ with $x_n = (an+b) y_n$. Then
    $\Pi_\mathrm{both}(\bfX)=\bfX$, where
    $\Pi_\mathrm{both}=\Pi_\mathrm{row}+\Pi_\mathrm{col}-\Pi_\mathrm{row}\circ \Pi_\mathrm{col}$.
\end{corollary}

\begin{remark}
    Note that multiplication of a series by $an+b$ means that the multiplicities of its
    characteristic roots increase by $1$.
\end{remark}

\begin{corollary}
    \label{cor:poly}
    Let $\Pi_\mathrm{row}$ be the projection on the row trajectory space of the
    polynomial of order $m$, $\Pi_\mathrm{col}$ be the projection on the column
    trajectory space of the polynomial of order $k$.  Then for the polynomial $\tX =
    P_{m+k+1}$ of order $m+k+1$ we have $\Pi_\mathrm{both}(\bfX)=\bfX$.
\end{corollary}

\subsection{Separability}
We expect that if  a time series component is governed by a minimal LRR and this LRR is known,
then the series component can be separated by a suitable version of SSA with projection better than it can be done by Basic SSA.

Using the notion of separability, we can formulate this improvement as follows.
Let $\tX=\tX^{(1)}+\tX^{(2)}$. We will say that a time series component
$\tX^{(1)}$ is separated by SSA with projection if $\tX^{(1)}=\bfC$, where $\bfC$ is as in \eqref{eq:ssaproj}.

Let $\tX^{(1)}$ be a series of finite rank, $\tX=\tX^{(1)}+\tX^{(2)}$.
Similar to  \cite{Golyandina.etal2001}, where conditions for separability by SSA with
centering are considered, the following conditions of separability can be obtained.

\begin{enumerate}
    \item Basic SSA:\\ $\tX^{(1)}$ and $\tX^{(2)}$ are separable if (if and only if, by
    definition) their row and column
    spaces are orthogonal.
    \item SSA with row projection on the row space of $\tX^{(1)}$: \\$\tX^{(1)}$
    and $\tX^{(2)}$ are separable if their row spaces are
    orthogonal.
    \item SSA with column projection on the column space of $\tX^{(1)}$:\\
    $\tX^{(1)}$ and $\tX^{(2)}$ are separable if their column spaces are
    orthogonal.
    \item SSA with double projection on the row and column space of $\tY$,
    where $\tX^{(1)}$ and $\tY$ are such that $x^{(1)}_n = (an+b) y_n$, $a\neq 0$:\\
    $\tX^{(1)}$ and $\tX^{(2)}$ are separable if the row and column spaces of
    $\tY$ and $\tX^{(2)}$ are orthogonal.
\end{enumerate}

Note that the separability by SSA with projection is always strong, since projections on linear spaces are uniquely defined.

For the approximate separability, where $\tX^{(1)} \approx \bfC$, the approximate orthogonality
is necessary. Also, the asymptotic separability
can be considered by analogy with the conventional separability
for Basic SSA and SSA with centering.

Recall that the usual double centering in SSA corresponds to  a constant
series $\tY$ and therefore to a linear series $\tX^{(1)}$. Orthogonality to a
constant series is a much weaker condition than that to a linear
series (moreover, the condition of orthogonality to a  linear series can never
be exactly satisfied). In particular,
any sinusoid with frequency $\omega$ is asymptotically separable from the linear trend and the exact separability by SSA with projection takes place
if $L\omega$ and $K\omega$ are integers, that is, if $L$ and $K$ are divisible by the period of the sinusoid.
Therefore, for extraction of linear trends, the double centering is recommended.

In the case of a polynomial trend of degree larger than $1$, the conditions of exact separability cannot be satisfied at all,
even for SSA with double projection.
However, we still can expect that in the case of polynomial trends,
SSA with double projection also will work better than SSA with only row or
column projections and also better than Basic SSA.

\subsection{Algorithm}

Let us summarize the steps of SSA with projection in the form of algorithms, splitting the whole algorithm into
decomposition and reconstruction.

\begin{algorithm}{SSA with projection: decomposition}{alg:proj_ssa_decomp}
    \Input{The time series $\tX$ of length $N$, the window length $L$, an orthonormal
    basis of the column projection space $(P_i, i=1,\ldots, p)$ and
    an orthonormal basis of the row projection space $(Q_i, i=1,\ldots, q)$.
    Either $p$ or $q$ can be zero.}

    \Output{Decomposition of the trajectory matrix on elementary
        matrices $\bfX=\bfX_1+\ldots+\bfX_d$, where $\bfX_i=\sqrt{\sigma_i} U_i V_i^\rmT$ are rank-one matrices.}

    \State Construct the trajectory matrix $\bfX = \cT_\mathrm{SSA} (\tX)$.
    \State Subtract the row projection: $\bfX'=\bfX-\bfC$, where
    \bea
    \bfC=\Pi_\mathrm{row}(\bfX)= \sum_{i=1}^q\sigma^{(r)}_i \wtilde{P}_i Q_i^\rmT,
    \eea
    $\sigma^{(r)}_i=\|\bfX Q_i\|$, $\wtilde{P}_i=\bfX Q_i/\sigma^{(r)}_i$.
    \State Subtract the column projection: $\bfX''=\bfX'-\bfC'$, where
    \bea
    \bfC'=\Pi_\mathrm{col}(\bfX')= \sum_{i=1}^p\sigma^{(c)}_i P_i \wtilde{Q}_i^\rmT,
    \eea
    $\sigma^{(c)}_i=\|\bfX'^\rmT P_i\|$, $\wtilde{Q}_i=\bfX'^\rmT P_i/\sigma^{(c)}_i$.
    \State Construct a  decomposition $\bfX''=\sum_{i=1}^{d''} \bfX''_i$, where
    $\bfX''_i=\sqrt{\lambda'_i} U''_i (V''_i)^\rmT$; it can be performed by
    Decomposition step of Basic  SSA.
    \State As a result, $\bfX=\sum_{i=1}^d \bfX_i$,
    where $d=p+q+d''$,
    $\bfX_{i}=\sigma^{(r)}_i \wtilde{P}_i Q_i^\rmT$ for $i=1,\ldots,q$,
    $\bfX_{i+q}=\sigma^{(c)}_i P_i \wtilde{Q}_i^\rmT$ for $i=1,\ldots,p$, and
    $\bfX_{i+p+q}=\sqrt{\lambda''_i} U''_i (V''_i)^\rmT$ for $i=1,\ldots,d''$.
\end{algorithm}

Similar to Basic SSA, SSA with projection provides a decomposition on matrices orthogonal by Frobenius; therefore,
contributions of $\bfX_i$ are given by $\|\bfX_i\|^2/\|\bfX\|^2$.
However, the obtained decomposition into a sum of rank-one matrices can be non-minimal (their number is larger than the rank of $\bfX$),
if at least one basis vector
used for the projections does not belong to the column (row) trajectory
space.

\begin{algorithm}{SSA with projection: reconstruction}{alg:basic_ssa_reconstr}
    \Input{Decomposition $\bfX=\bfX_1+\ldots+\bfX_d$ and
    grouping $\{1,\ldots,d\}=\bigsqcup_{j=1}^m I_j$, which does not split
        the first $p+q$ projection components, where $q$ and $p$ are the numbers of row and column
        projection components.
        }
    \Output{Decomposition of time series on identifiable components
        $\tX=\tX_1+\ldots+\tX_m$.}

    \State Construct the grouped matrix decomposition
    $\bfX=\bfX_{I_1}+\ldots+\bfX_{I_m}$, where $\bfX_I=\sum_{i\in I} \bfX_i$.
    \State Compute $\tX=\tX_1+\ldots+\tX_m$, where $\tX_i=\cT^{-1} \calH (\bfX_{I_i})$.
\end{algorithm}

The only essential difference with the reconstruction by Basic SSA  is that
the set of the matrices $\bfX_i$, $i=1,\ldots,p+q$, produced by
projections, should be included in the same group.

Note that formally, the sets $\{P_i, i=1,\ldots, p\}$ and
$\{Q_i, i=1,\ldots, q\}$ can be arbitrary.  However, if the model of the series is partly known, then in the context of SSA
this means that a time series component satisfies an LRR and we know its characteristic roots (see
Section~\ref{sec:lrr}). Therefore, to extract, for example, a sine wave using projections, we should
know its period, and to extract an exponential trend, we should know its rate.  These conditions are
often  too restrictive.  A clear exception is extraction of polynomial trends of a degree $m$,
when there is the unique characteristic root equal to $1$ of multiplicity $m+1$ and we should assume only the degree of the polynomial trend to obtain its trajectory space.

\section{Examples}
\label{sec:ex}

The presented examples are related to finding polynomial trends.
For convenience, if the row and column projections are performed on the subspace generated by polynomials of degree $q-1$
and $p-1$ respectively, then we denote the method as ProjSSA($q$,$p$).
Recall (see Corollary~\ref{cor:poly}) that the choice ProjSSA($q$,$p$)
corresponds to extraction of a polynomial trend of degree $q+p-1$. In ProjSSA($q$,$p$),
the projection part of the decomposition, i.e., the matrix $\bfC$ in \eqref{eq:ssaproj}, consists of
$p+q$ rank-one matrices.
ProjSSA(1,1) is used for extraction of a linear trend. The zero value for $p$ or $q$ means that the corresponding projection
is not performed.

All the examples are implemented in \pkg{R} \citep{R2015} with the help of the \pkg{Rssa} package \citep{Korobeynikov.etal2015}.
For example, to perform ProjSSA($q$,$p$) for a time series taken from the variable  \code{x} with a window length \code{L},
the following code should be called:
\begin{Code}
s <- ssa(x, L = L,
         row.projector = q,
         column.projector = p)
r <- reconstruct(s,
        groups = list(trend = 1:nspecial(s)))
plot(r, add.residuals = FALSE,
     plot.method = "xyplot", superpose = TRUE)
\end{Code}
For more details on \pkg{Rssa}, see the help files in \cite{Korobeynikov.etal2015}.

\subsection{SSA with projection and regression}
\label{sec:gasoline}
Let us demonstrate that the conventional linear regression and SSA with double centering,
i.e., ProjSSA(1,1), use different statements of the solved problem and therefore can yield different results.
It is clearly seen in short time series. For long time series the results are very close.
Also, in the model of linear regression with Gaussian noise, the regression solution is optimal.
Therefore, to demonstrate the difference, we consider a time series, which contains a seasonal component.

Here we examine the time series `Gasoline' taken from \cite{Abraham.Ledolter1983} and containing the data
GASOLINE DEMAND, MONTHLY, Jan 1960 -- Jun 1967, ONTARIO, GALLON MILLIONS. 

\bfgh
        \begin{center}
        \includegraphics[width=3.5in]{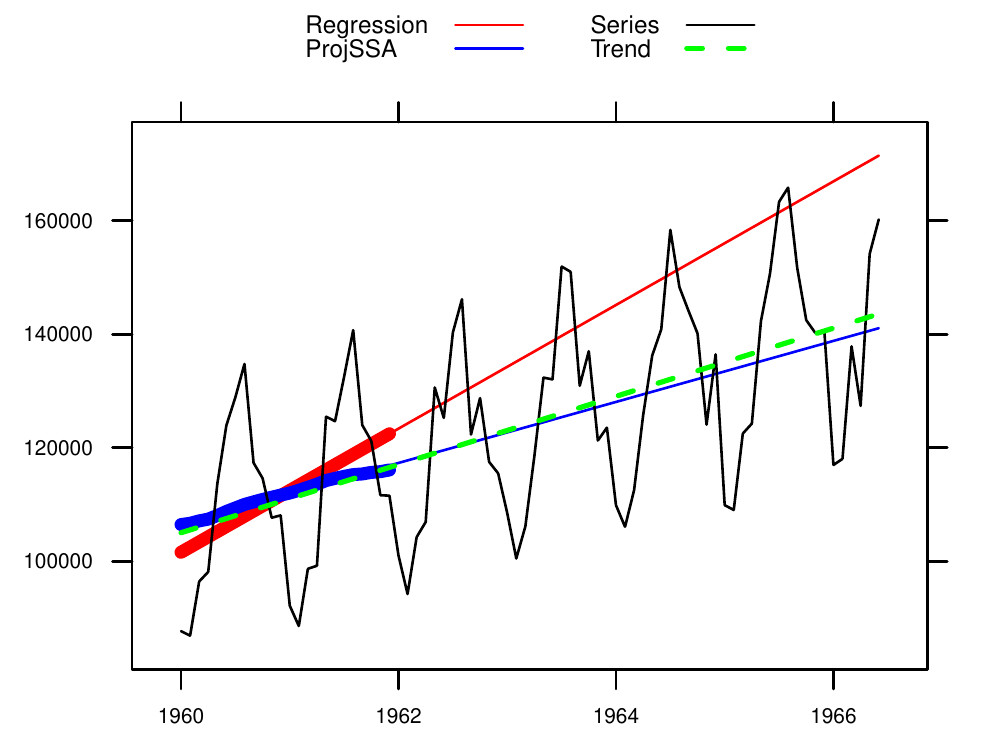}
        \end{center}
        \caption{`Gasoline': SSA with projection, linear trend detection.}
        \label{fig:gasoline_centering}
\efg

Let us consider the first two years and apply the linear regression and ProjSSA(1,1) with $L=12$.
To show the difference, we continue the linear regression line with the help of the estimated coefficients.
In the \pkg{Rssa}, a method of forecasting for SSA with projection is implemented.
Since it is not proved yet, we will construct the forecast by a linear regression
applied to the reconstruction, which is performed by ProjSSA(1,1).
Note that the forecasting procedure from \pkg{Rssa} provides a similar prediction.
As a benchmark, the linear regression constructed by the whole series is considered.

One can see in Figure~\ref{fig:gasoline_centering} that the ProjSSA(1,1) linear trend (blue) is very close
to a linear trend constructed by the whole long time series (green).
The linear regression line (red) gives a much worse approximation of the trend.
This is explained by the following reasons. The least-squares approach to the linear regression estimation minimizes the prediction error
and therefore the seasonal component can shift the linear regression trend.
For ProjSSA(1,1), the seasonal component is well separated from the linear trend,
since for the chosen parameters $L=K=12$ are divisible by the seasonal period $12$.

\subsection{SSA with projection and Basic SSA}
\label{sec:co2}
The example introduced in this section demonstrates that both SSA with projection and Basic SSA
can extract trends in a similar manner.
Let us consider the example `co2' (Mauna Loa Atmospheric $\mathrm{CO}_2$ Concentration, 468 observations, monthly from 1959
to 1997 \citep{Keeling1997}).

\bfgh
        \begin{center}
        \includegraphics[width=3.5in]{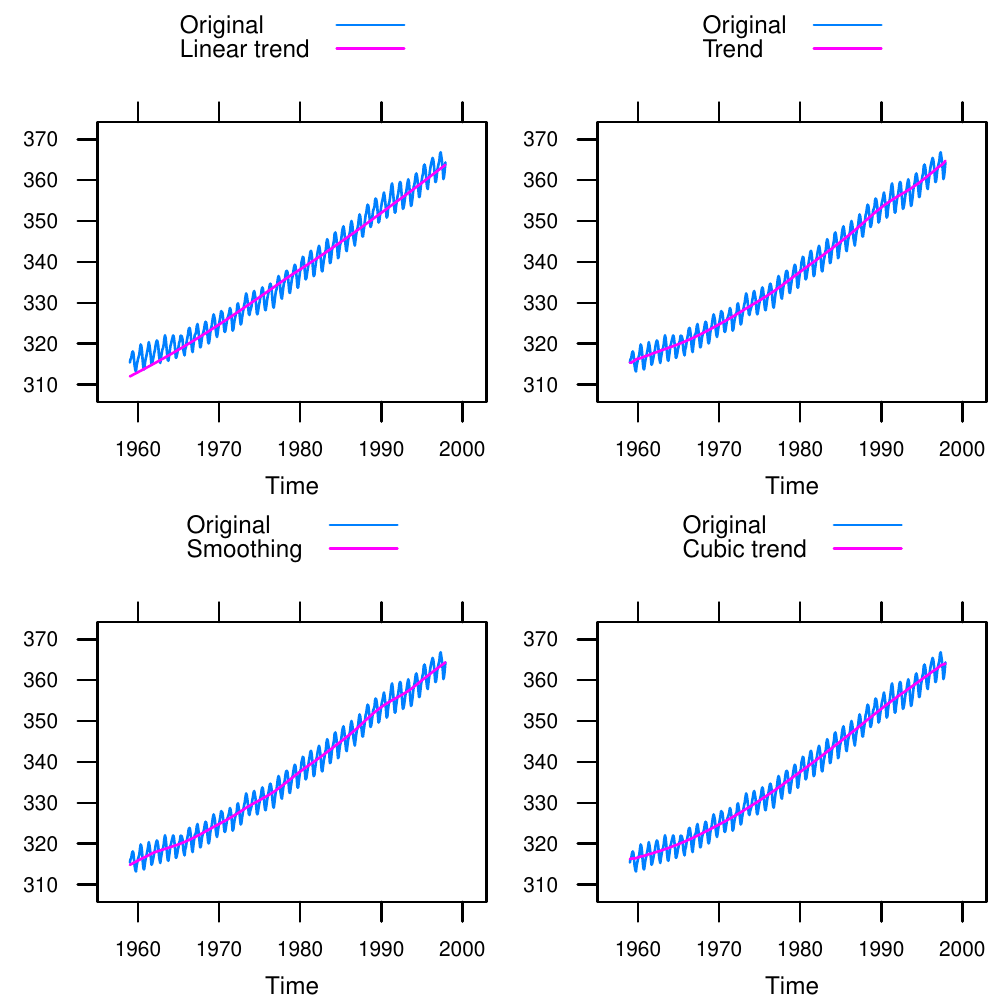}
        \end{center}
        \caption{co2: Reconstructions of the trend. Left-top: ProjSSA(1,1), $L=228$; right-top: ProjSSA(1,1), $L=228$, complemented by the ET $5$ and $8$;
        left-bottom: ProjSSA(1,1), $L=36$; right-bottom: ProjSSA(2,2), $L=228$.}
        \label{fig:pssa_rec_all}
\efg

We start with extraction of the linear trend and therefore choose
ProjSSA(1,1) to perform SSA with double centering.

By analogy with SSA, large window lengths help to extract separable series components,
while small window lengths correspond to smoothing.
Therefore, we take $L=228$, which is divisible by $12$ and is close to half of the time series length to obtain better separability,
and  a small value $L=36$ to smooth the series.
Three of four versions of the extracted trends presented in Figure~\ref{fig:pssa_rec_all} almost coincide.

For the choice $L=228$, the extracted trend is close to linear, see Figure~\ref{fig:pssa_rec_all} (left-top).
Certainly, the accurate trend of `co2' series is not linear.
However, the projection components can be supplemented by the 1st and 4th SVD components (ET5,8) to improve the trend
(Figure~\ref{fig:pssa_rec_all} (right-top)).
Figure~\ref{fig:pssa_rec_all} (left-bottom) shows the result of smoothing with $L=36$. Finally,
the result of ProjSSA(2,2) with $L=228$, which is designed for extraction of a cubic trend, is depicted in Figure~\ref{fig:pssa_rec_all} (right-bottom).
The extracted trend is very similar to that in \cite{Golyandina.Korobeynikov2013},
which was extracted by Basic SSA (not depicted).

Identification of the components in the decomposition produced by SSA with projection
is exactly the same as it is performed in Basic SSA.

\subsection{Numerical comparison}
\label{sec:comparison}

The real-life examples presented in Sections~\ref{sec:gasoline} and \ref{sec:co2}
show that the results of Basic SSA, SSA with projection and linear regression
can be either different or similar. To understand, what method is better, let us perform a numerical study.

We consider a time series of length $N=199$ with the common term
\begin{equation}
 x_n=t_n+s_n+\ve_n,
\end{equation}
where $t_n$ is a trend, $s_n=A\sin(2\pi n \omega +\phi)$, $\ve_n$ is a Gaussian white noise with standard deviation
$\sigma$.

For obtained estimations $\hat{t}_n^{(i)}$, where $i$ is the number of series with $i$th realization of noise $\ve_n^{(i)}$, $i=1,\ldots,M$,
we will calculate the root-mean-square error (RMSE)
as $\sqrt{\frac{1}{MN}\sum_{i=1}^M \sum_{n=1}^N(\hat{t}_n^{(i)}-t_n)^2}$.

\smallskip
\emph{Linear trend and sine wave.}
Let us start with the noiseless case with $\sigma=0$ and therefore take $M=1$.
Let $t_n=a n+b$. We fix $a=1$, $b=-100$, $A=1$ and change $\omega$ from $0.02$ to $0.1$ (that is,
the period is changed from $50$ to $10$).

Since the result of the least-square method strongly depends on the form of the residual, we consider
the values of the phase, $\phi=0$ and $\phi=\pi/2$.

Figure~\ref{fig:sim_sin} (left) contains the RMSE values in the case $\phi=0$ for Basic SSA with reconstruction by ET1--2, ProjSSA(2,0),
ProjSSA(1,1) with $L=100$, and for the linear regression.  One can see that the worse cases for ProjSSA(1,1)
are approximately equal to the best cases for the linear regression. 

\bfgh
        \begin{center}
        \includegraphics[width=1.75in]{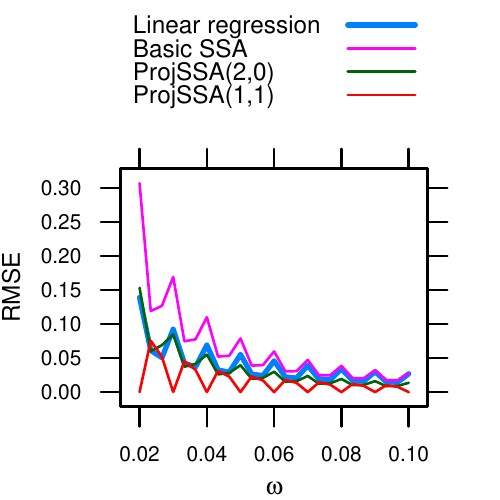}
        \includegraphics[width=1.75in]{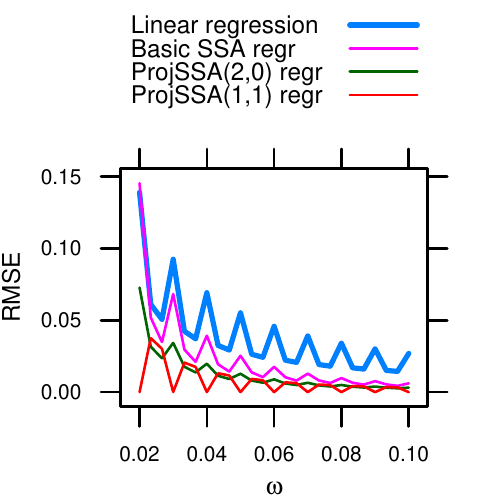}
        \end{center}
        \caption{Dependence of the RMSE of linear-trend estimates on frequency of the periodic component. $\phi=0$.}
        \label{fig:sim_sin}
\efg

\bfgh
        \begin{center}
        \includegraphics[width=1.75in]{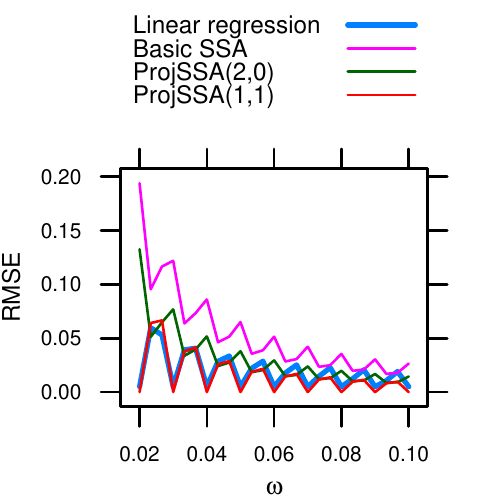}
        \includegraphics[width=1.75in]{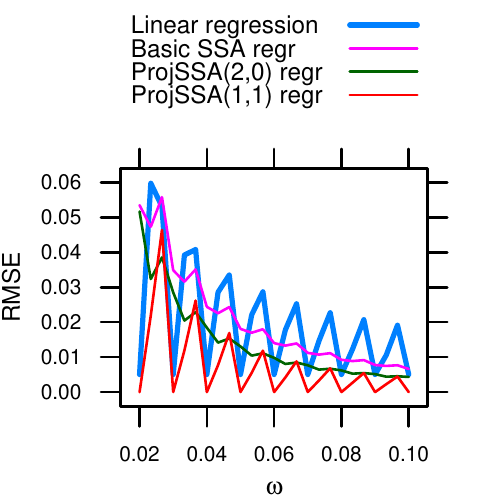}
        \end{center}
        \caption{Dependence of the RMSE of linear-trend estimates on frequency of the periodic component, $\phi=\pi/2$.}
        \label{fig:sim_cos}
\efg

In Section~\ref{sec:gasoline}, we perform forecasting by the linear regression applied to the trend reconstruction.
Figure~\ref{fig:sim_sin} (right) contains the RMSE for the linear regression lines constructed in this way; `regr' is added to the legend. The ordering of the SSA methods is generally the same, while the SSA methods become better than the linear regression.
Probably, $0$ is one of the worst values of $\phi$ for linear regression.

Now consider $\phi=\pi/2$ as  one of the best cases for the linear regression.
The behavior of the errors is quite different (Figure~\ref{fig:sim_cos} (left)).  However,
the accuracy of ProjSSA(1,1) is still better than that of the linear regression.
Linear least-square approximation of the SSA reconstructions considerably improves the
accuracy of the SSA methods (Figure~\ref{fig:sim_cos} (right)).

Note that zero values of the RMSE for ProjSSA(1,1) for frequencies $\omega=0.01k$ are explained by the theory,
since then $L\omega$ and $K\omega$ are integers.
The errors for ProjSSA(2,0) lie between that for Basic SSA and ProjSSA(1,1).
It is interesting that the minimal errors for Basic SSA are achieved
for the middle points, when $L\omega+0.5$ and $K\omega+0.5$ are integers.

\smallskip
\emph{Cubic trend and sine wave.}
Let us consider a more complex case of the cubic trend $t_n=0.0001 n^3$.
Since there is no exact separability for any choice of parameters,
the results are unpredictable.
Figures~\ref{fig:sim_cubic_sin} (left) and \ref{fig:sim_cubic_cos} (left) contain the RMSE values for Basic SSA with reconstruction by ET1--4, ProjSSA(4,0),
ProjSSA(2,2) with $L=100$ and for the cubic regression.
One can see that  ProjSSA(2,2) is the best method for $\phi=0$, while it is just comparable with the linear regression for $\phi=\pi/2$.
Note that here the best parameters for ProjSSA(2,2) do not correspond to the case when $L\omega$ and $K\omega$ are integers.
The cubic least-square approximation of the reconstructed trend again improves the estimates
(Figures~\ref{fig:sim_cubic_sin} (right) and \ref{fig:sim_cubic_cos} (right)).

\bfgh
        \begin{center}
        \includegraphics[width=1.75in]{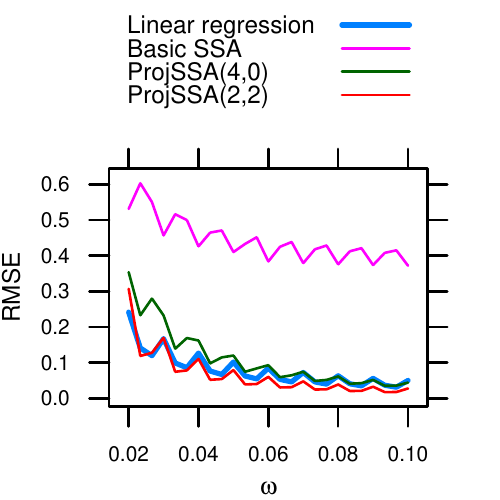}
        \includegraphics[width=1.75in]{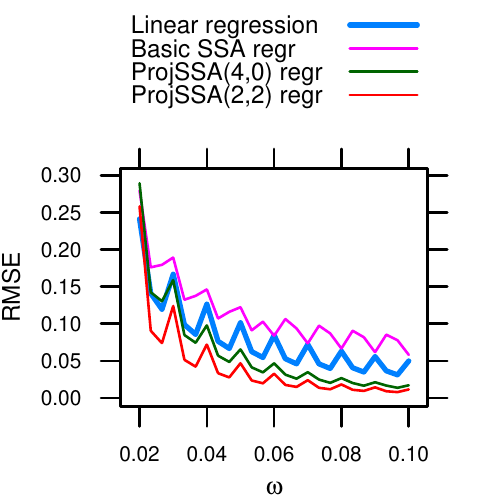}
        \end{center}
        \caption{Dependence of the RMSE of cubic-trend estimates on frequency of the periodic component, $\phi=0$.}
        \label{fig:sim_cubic_sin}
\efg

\bfgh
        \begin{center}
        \includegraphics[width=1.75in]{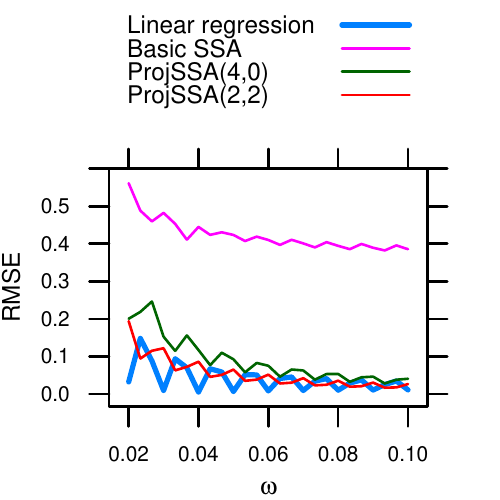}
        \includegraphics[width=1.75in]{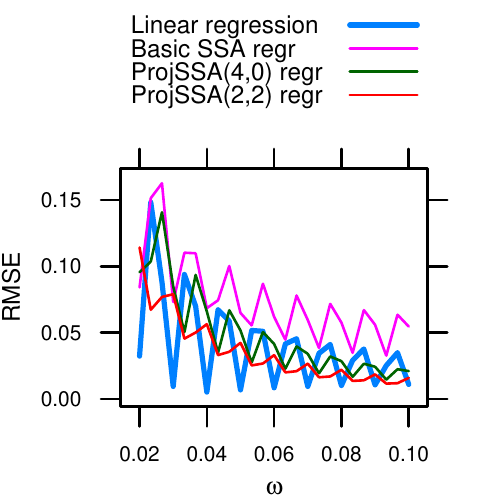}
        \end{center}
        \caption{Dependence of the RMSE of cubic-trend estimates on frequency of the periodic component, $\phi=\pi/2$.}
        \label{fig:sim_cubic_cos}
\efg

Basic SSA fails for the chosen parameters because of lack of strong separability: the fourth trend component has
a contribution comparable with the contribution of the periodic components that causes their mixture.

Note that one of the modifications described in \cite{Golyandina.Shlemov2014}, Iterative O-SSA,  can be used
to get strong exact separability for the considered noiseless examples. However,
we do not involve this modification into the comparison, since Iterative O-SSA is not able to remove noise
and should be applied after denoising in nested manner, while the compared methods are able to extract the trend without denoising.

\smallskip
\emph{Linear trend and noise.}
For the data which satisfy the model of the linear regression with white Gaussian noise,
that is, for the amplitude $A$ equal to zero, we take $\sigma=1$ and use $M=1000$.
As expected, the smallest error $0.10$ is achieved for the regression estimate. However,
the RMSE of the ProjSSA(1,1) estimate equal to $0.12$ is very close to $0.10$.
The error of the Basic SSA is equal to $0.17$.
Application of linear regression to the results of SSA reconstruction
improves the SSA estimates. The RMSE for ProjSSA(1,1) and Basic SSA become
equal to $0.115$ and $0.104$ respectively.

We do not show the results when the series has both periodic component and noise,
since the errors are intermediate. To keep the advantage of SSA with projection, the noise standard $\sigma$ should be
considerably smaller than the amplitude $A$ of the periodic component.

\section{Conclusion}
\label{sec:concl}

The considered combination of singular spectrum analysis, which does not need a series model given in advance,
and of a subspace-based parametric approach, which is incorporated by means of projections to subspaces given in advance, proves successful for extraction of polynomial (especially, linear)
trends, when the residual has unknown structure and can include deterministic oscillations, e.g., the seasonality.

The general form of projections of columns and rows of the trajectory matrix, which keeps
this trajectory matrix, was obtained. It was proved that projections to the row and column subspaces (so-called double projection)
of the trajectory matrix of a series $\tY$ are related to extraction of the series $(an+b)\tY$. In particular, the linear trend
can be obtained by double projection to the column and row subspaces of a constant series.
The formulated conditions of separability of a series component, which is kept by projections, show that
if a series component can be represented in the form $(an+b)\tY$, then the double projection is preferable.

Thus, the theory provides an additional theoretical support to  SSA with double centering (ProjSSA(1,1)), which was known before,
and also enlarges the range of applications of semi-nonparametric modifications of Basic SSA.

\smallskip
Applications of SSA with projection considered in the paper were related to the extraction of
a polynomial trend, since its trajectory space is determined by the polynomial degree only.

We showed on the example `Gasoline' that the linear regression approach can be inadequate for short series and
large oscillations, in comparison with ProjSSA(1,1).
Comparison of different SSA versions applied to the `co2' data demonstrates that even if the model of a series component
used for projection is wrong, the non-parametric part of SSA with projection can correct the bias.

A numerical study was performed for a better understanding of the difference between SSA with projection
and the linear regression approach.
First, it appears that if we extract a polynomial trend by SSA with projection, then
the polynomial least-squares approximation of the trend reconstruction can considerably improve the
accuracy.

The second found effect is related to the influence of the residual geometry on the estimate accuracy.
In the considered example, we changed the phase of a sinusoid.
The SSA estimates slightly depend  on the phase, while the regression estimates demonstrate a considerable dependence.

Numerical experiments confirm that for a linear trend and a sine wave residual, ProjSSA(1,1) is more accurate than the linear regression estimate.
For a noisy linear trend, when the model of the linear regression if fulfilled, the linear regression estimate is slightly more accurate than SSA.
Thus, we can formulate conditions, when SSA with double projection can be recommended for use:
series has a linear or polynomial trend (the polynomial degree is not large) and the regular
oscillations are considerably larger than the noise level.

The further investigation can be performed in two directions. First, the forecasting algorithm for ProjSSA($m$,$k$) implemented in \pkg{Rssa} should be proved.
Then, the idea to use projection to
involve the structure of a supporting series looks promising.

\bibliographystyle{plain}

\end{document}